  \documentstyle[prl,aps,twoside,floats,epsf]{revtex}
  \begin{document}
  \draft
  \twocolumn[\hsize\textwidth\columnwidth\hsize\csname
  @twocolumnfalse\endcsname
  \draft 
  \title{Theory of optical conductivity in doped manganites}
  \author{ A.S. Alexandrov$^{1,*}$ and A.M. Bratkovsky$^{2,\dagger}$}
  \address{$^{1}$Department of Physics, Loughborough University,
  Loughborough LE11 3TU, UK\\
  $^{2}$Hewlett-Packard Laboratories, 3500~Deer~Creek
  Road, Palo Alto, California 94304-1392 }
  \date{January 26, 1999 }
  \maketitle
  \begin{abstract}
  The frequency and temperature
  dependence of the optical conductivity of
  ferromagnetic manganites
  is explained within the framework of the bipolaron theory.
  As these materials are cooled below the Curie temperature, the
  colossal magnetoressitance (CMR) is accompanied by a massive transfer
  of the spectral weight of the optical conductivity to lower
  frequencies. As with the CMR itself, this change in the optical
  conductivity is explained by the dissociation of bipolarons 
 into small polarons by exchange interaction with the
  localized Mn spins during the transition to the low temperature
  ferromagnetic phase.
  \end{abstract}
  \pacs{71.30.+h, 71.38.+i, 72.20.Jv, 78.20.Bh}
  \vskip 2pc ] 
  \narrowtext
  It is well established that carriers in manganites
  are strongly coupled with the lattice vibrations \cite{mil,bis}.
  As we have recently shown \cite{alebra} the interplay of the
  electron-phonon
  and
  exchange interactions results in a
  current carrier density collapse (CCDC) at the transition.
  Owing to the strong electron-phonon interaction, polaronic carriers
  are bound into almost immobile bipolarons
  in the paramagnetic phase. A few thermally excited
  non-degenerate polarons
  polarize localized Mn $d$ electrons. As a result, the exchange
  interaction breaks
  bipolarons below $T_c$ if the
  $p-d$ exchange energy $J_{pd}S$
  of the polaronic carriers with the localized Mn $d$ electrons
  is larger than the bipolaron binding energy
  $\Delta$. Hence, the density of current carriers (polarons) suddenly
  increases
  below $T_c$, which explains the resistivity peak and CMR
  experimentally observed in many ferromagnetic oxides
  \cite{van,helmolt,jin}.
  We have also shown \cite{alebra2} that
  CCDC
  explains the giant isotope effect \cite{mul}, the tunneling gap
  \cite{tun},
  the specific heat anomaly \cite{ram}, along with the temperature
  dependence of the dc resistivity
  \cite{sch}.

  Recent studies of the optical conductivity \cite{oki1,oki2,kim,ish} and
  photoemission
  \cite{des} unambiguously
  confirmed a non-metallic origin of the ferromagnetic
  phase.
  In particular,
  a broad incoherent spectral feature \cite{oki1,oki2,kim,ish} in the
  midinfrared region
  and a pseudogap
  in the excitation spectrum \cite{des} were observed, while the
  coherent Drude weight
  appeared to be {\em two orders} of magnitude smaller \cite{oki2,kim} than
  expected for a metal, or almost {\em absent}\cite{ish}.
  These and other studies \cite{gud} prove that carriers
  retain their polaronic
  character well below $T_c$, in agreement with our theory of
  CMR \cite{alebra}. However,
  a conspicuous sudden spectral weight transfer
  with temperature \cite{oki2,kim,ish} as well
  as a pronounced peak structure \cite{ish} in the optical
  conductivity below $T_c$ remain to be
  understood.
 
 In this Letter we propose a theory
  of the optical conductivity, including the {\em sudden spectral weight
  transfer} below the ferromagnetic transition,
  based on the idea of
  the current carrier density collapse. We show that the high-temperature
  optical conductivity is well described by the small bipolaron
  absorption, while
  the low temperature midinfrared band is due to absorption by small
  polarons.
  The magnetic bipolaron breaking below $T_c$ shifts the spectral weight
  from the bipolaronic peak to the polaronic one. We
  describe the optical spectra of the layered ferromagnetic ($T_c=125$K)
  crystals
  La$_{2-2x}$Sr$_{1+2x}$Mn$_2$O$_7$ \cite{ish} in the entire frequency and
  temperature
  range, and show that the optical data provides a strong evidence for CCDC.

  The optical intraband conductivity of a charge-transfer doped insulator
  with
  (bi)polaronic carriers
  is the sum of the polaron $\sigma_{p}(\nu)$ and bipolaron
  $\sigma_{b}(\nu)$ contributions
  at the given frequency $\nu$.
  Their frequency dependences are described in
  the literature \cite{bri,mah,alemot}. Both have almost a Gaussian
  shape given by
  \begin{eqnarray}
  \sigma_{\rm intra}(\nu)&=&{\sigma_{0}{\cal T}^{2}\over{\nu}}\Biggl[{n\over{
  \gamma_{p}}}
  \exp\left[-(\nu-\nu_{p})^{2}/\gamma_p^2\right]\nonumber\\
  &+&{x-n\over{\gamma_{b}}}
  \exp\left[-(\nu-\nu_b)^2/\gamma_b^2\right]\Biggr],
  \end{eqnarray}
  where $\sigma_{0}= 2\pi^{1/2}e^{2}/a$ is a constant with $a$ the lattice
  spacing,
  ${\cal T}$ the hopping integral, $n$ the (atomic) polaron density, and $x$
  the doping level.
  Here and further we take $\hbar=c=1$.

 Within the Holstein model with a
  local electron-phonon interaction, the
  polaron absorption has a maximum around
  $\nu_{p}=2E_{p}$\cite{kli,eag,rei},
  while the on-site bipolaron absorption has a maximum around
  $\nu_{b}=4E_{p}-U$ \cite{bri2}.
  Here $E_{p}$ is the polaronic (Franck-Condon) shift of the
  electron level and $U$ is the on-site (Hubbard) repulsion. The broadening
  of
  the absorption lines
  is due to the zero-point quantum fluctuations of ions,
  $\gamma_p=\gamma_b=(4E_{p}\omega)^{1/2}$,
  if the temperature is well below the
  characteristic phonon frequency $\omega$.
  The spectral shape, Eq.~(1), is applied in a wide frequency
  range, $\nu \gg\omega$, where the Franck-Condon principle applies
  (see, for example, \cite{mah}).

  The Holstein model with on-site bipolarons
  is highly unrealistic because of a very large on-site Coulomb repulsion
  and the long-range (Fr\"ohlich) electron-phonon interaction,
  which dominates in ionic
  solids. The latter is not reduced to a short-range interaction
  because heavy polarons cannot screen
  high-frequency crystal field oscillations in the low mobility solids.
  On the contrary, small polarons and small
  $inter$site bipolarons formed by the Fr\"ohlich interaction together
  with the deformation potential are rather
  feasible \cite{ale,alekor}.
  Applying the Franck-Condon principle in the adiabatic regime,
  $\nu\gg\omega$,
  one can readily generalize the (bi)polaronic absorption, Eq.~(1),
  to describe the optical conductivity of these quasi-particles, Fig.~1.
  \begin{figure}[t]
  \epsfxsize=2.4in
  \epsffile{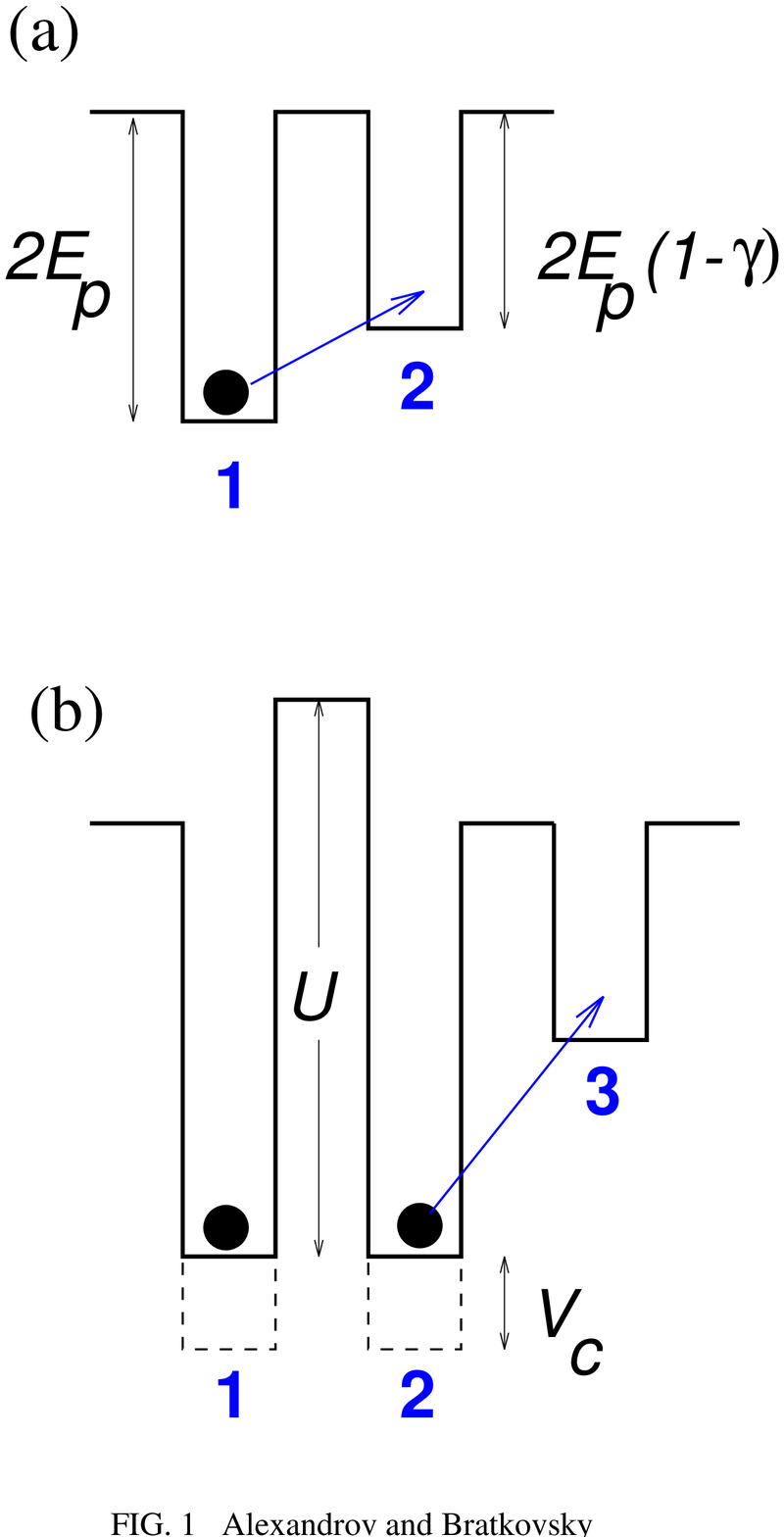 }
  \vspace{.2in}
  \caption{Adiabatic energy levels for the small polaron (a)
  and the inter-site small bipolaron (b). For notations see text.
  \label{fig:1}
  }
  \end{figure}
  The electron `sitting' on a site ``1'', Fig.~1(a), lowers its energy
  by an amount
  $2E_{p}$, with respect to an atomic level in the undeformed lattice,
  owing to the lattice deformation. If the electron-phonon interaction has a
  finite radius,
  the electron also creates some deformation around a neighboring site ``2'',
  lowering
  its energy level by an amount $2E_{p}(1-\gamma)$, where \cite{ale}:
  \begin{equation}
  \gamma=\sum_{\bf q}|\gamma({\bf q})|^2\left[1-\cos({\bf q \cdot a})\right]
  /\sum_{\bf q}|\gamma({\bf q})|^2
  \end{equation}
  with ${\bf a}$ the lattice vector connecting the neighboring sites.
  The coefficient $\gamma$ strongly
  depends on the radius of the interaction. In the Holstein model with ${\bf
  q}$-independent electron-phonon coupling, $\gamma({\bf q})$, this
  coefficient equals unity. Hence,
  there is
  no lattice deformation at the neighboring site. On the contrary, in the
  Fr\"ohlich case,
  $\gamma({\bf q})\propto 1/q$, and the coefficient is quite small, $\gamma
  \approx
  0.2-0.4$ \cite{ale}
  depending on the dimensionality of the system and the unit cell geometry.
  In that case, there is a significant lowering of the neighboring energy
  level
  and, as a result, of the polaron mass \cite{alekor}.
  Hence, generally, the peak energy in the polaron absorption
  is found at
  \begin{equation}
  \nu_{p}= 2 \gamma E_p,
  \end{equation}
  and the activation energy of the high-temperature dc-conductivity
  is $E_a=\gamma E_p/2$\cite{mah}. One can apply the same `frozen lattice
  distortion'
  arguments
  to the inter-site bipolaron absorption, Fig.~1b.
  The electron energy on a site ``2'' is $-2E_{p}
  -2E_{p}(1-\gamma)+ V_{c}$, where the first contribution is due to the
  lattice
  deformation
  created by the electron itself, while the second contribution is due to
  the
  lattice
  deformation around the site ``2'' created by the other electron
  of the pair on the site
  ``1'',
  which is the polaron-polaron attraction \cite{alemot}. After absorbing the
  quantum of radiation, the electron hops from site ``2''
  to the empty site ``3''
  into a state with the energy $-2E_{p}(1-\gamma)$,
  which corresponds to an absorption frequency
  \begin{equation}
  \nu_{b}= 2E_{p}-V_{c},
  \label{eq:nub}
  \end{equation}
  where now $V_{c}$ is the inter-site Coulomb
  repulsion.
  The quantum broadening of the polaronic and bipolaronic absorption
  is given by $\gamma_p=\gamma_b= (4\gamma E_p\omega)^{1/2}$. Since
doped   manganites are intrinsically disordered, their dielectric
  properties are 
inhomogeneous, and so is $E_p$, which fluctuates with a characteristic
impurity broadening $\Gamma_{\rm im}$. The convolution of the polaronic and
bipolaronic absorption lines with the Gaussian distribution of $E_p$
results in their having different linewidths, 
$\gamma_p= 2(\gamma E_p\omega +\gamma^2\Gamma_{\rm im}^2 )^{1/2}$ and
$\gamma_b=2(\gamma E_p\omega+\Gamma_{\rm im}^2 )^{1/2}$
for polaronic and bipolaronic
absorption, respectively.
  The Coulomb repulsion $V_c$ can be readily estimated as
  $ V_{c}=2E_p - \nu_b$ from (\ref{eq:nub}).
 
  The total absorption is the sum of the intraband polaronic and
  bipolaronic terms, Eq.~(1),
  and the interband absorption,
  $\sigma(\nu)=\sigma_{\rm intra}(\nu)+\sigma_{\rm inter}(\nu)$.
  In the layered compounds like La$_{2-2x}$Sr$_{1+2x}$Mn$_2$O$_7$, the
  intraband contribution
  to the out-of-plane conductivity is negligible \cite{ish}. Hence, one can
  take the c-axis
  optical conductivity $\sigma_{c}(\nu)$
  as a measure of the interband contribution to the in-plane
  conductivity with a scaling factor, $s$, $\sigma_{\rm inter}(\nu)\simeq
  s\sigma_{c}(\nu)$.
  The scaling factor $s$ is the square of the ratio
  of the in-plane components of the dipole matrix element
  for the interband transitions to its $z$ component ($z$ is the
  out-of-plane
  direction).
  It can be readily determined by comparing the in-plane and out-of-plane
  optical
  conductivities at high frequencies, where intraband absorption is
  irrelevant.
  The result of the comparison of the present theory with the experiment
  \cite{ish} is shown
  in Fig.~2. 
  \begin{figure}[t]
  \epsfxsize=3.4in
  \epsffile{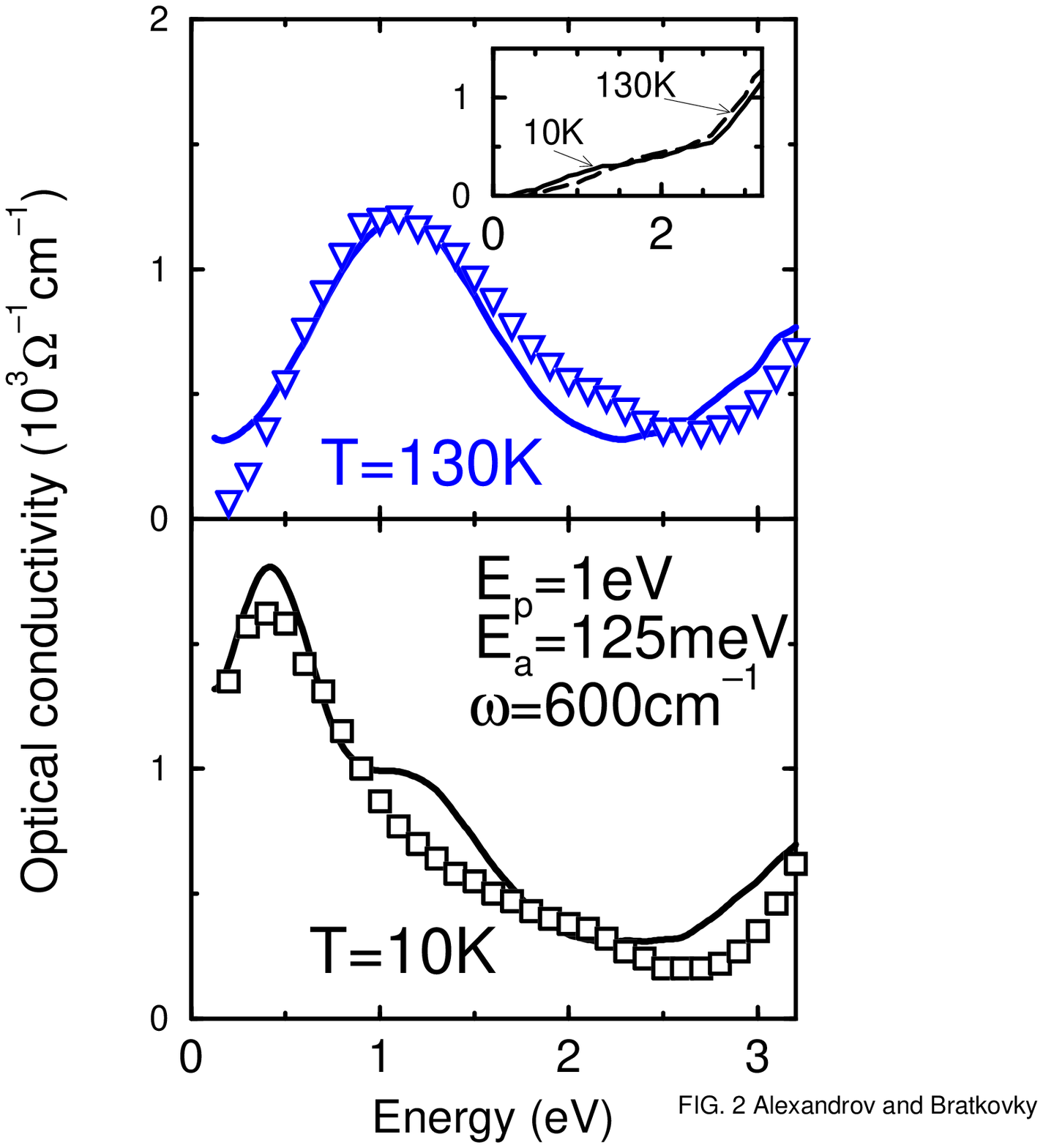 }
  \vspace{.2in}
  \caption{Optical conductivity of La$_{2-2x}$Sr$_{1+2x}$Mn$_2$O$_7$
  [15] compared with the theory (solid line) above $T_{c}$ (top panel) and well below
  $T_{c}$ (bottom panel). Inset: c-axis optical conductivity.
  \label{fig:2}
  }
  \end{figure}
At temperatures above the transition ($T=130$K) the  polaron
  density
  is very low
  owing to  CCDC \cite{alebra}, so the intraband conductivity is due to
  bipolarons only,
  \begin{equation}
  \sigma(\nu)={\sigma_{0} x{\cal T}^{2}\over{\nu \gamma_b}}
  \exp\left[-(\nu-\nu_b)^{2}/\gamma_b^2\right] +s \sigma_c(\nu).
  \end{equation}
  This expression fits the experiment fairly well with $\nu_{b}~=~1.24$ eV
  and
  $\gamma_{b}~=~0.6$ eV, Fig.~2.
  The scaling factor is estimated as $s=0.6$. When the temperature drops
  below $T_c$, at least some of the
  bipolarons break apart by the exchange interaction with Mn sites, because
  one of the spin-polarized polaron bands
  falls suddenly below the bipolaron level by an amount
  $(J_{pd}S-\Delta)/2$,
  Fig.~3 \cite{alebra}.
  \begin{figure}[t]
  \epsfxsize=3.4in
  \vspace{.2in}
  \epsffile{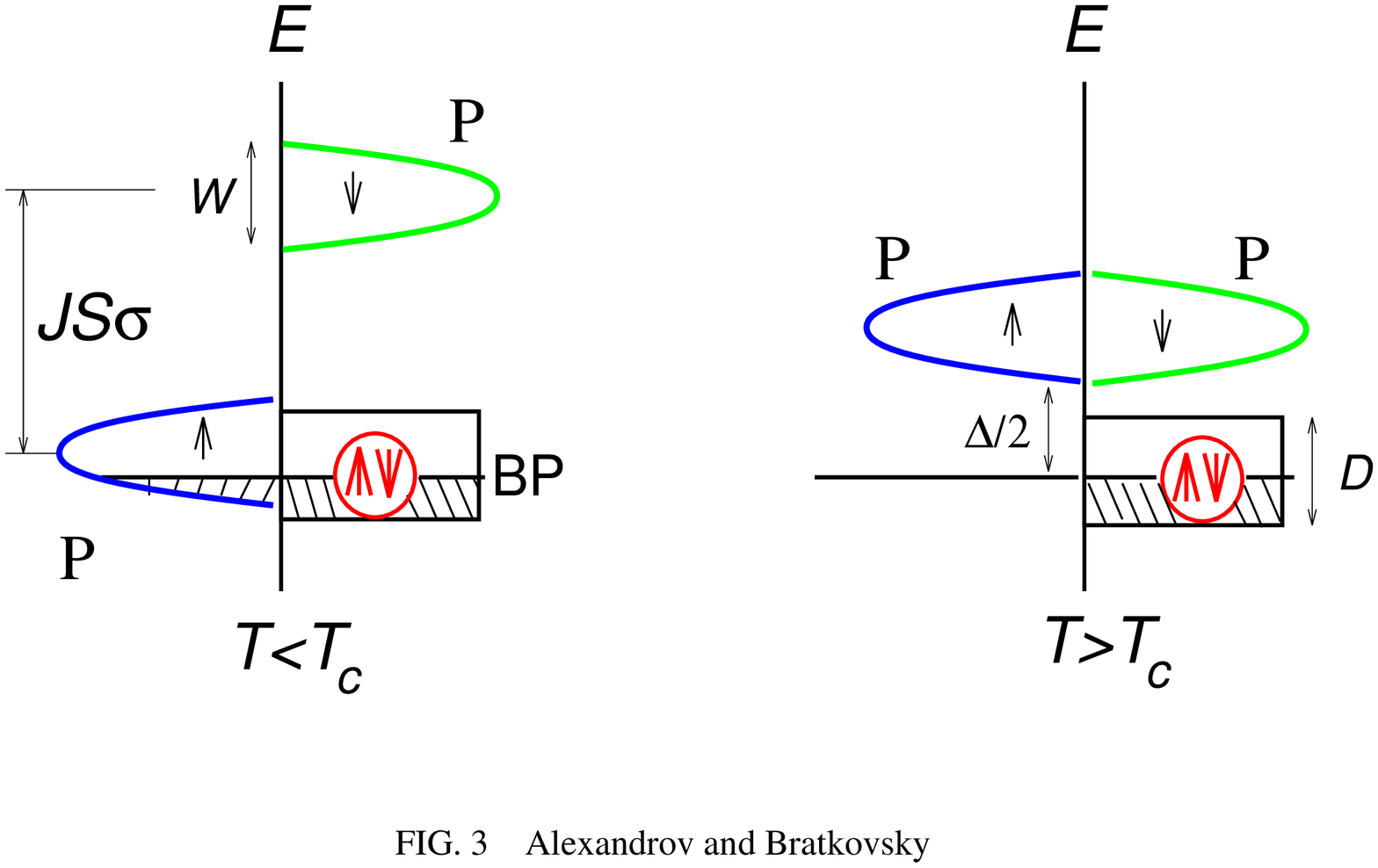}
  \vspace{.2in}
  \caption{Spin-polarized polaron band in the ferromagnetic phase ($T<T_c$)
  overlaps with the bipolaron (impurity) band, breaking up
  a fraction of the bipolarons.
  \label{fig:3}
  }
  \end{figure}
  The intraband
  optical conductivity is determined now by both the polaronic and
  bipolaronic
  contributions, Eq.~(1),
  and that explains the sudden spectral weight transfer from $\nu=\nu_{b}$
  to
  $\nu=\nu_{p}$
  observed below $T_{c}$ in the ferromagnetic manganites
  \cite{oki2,kim,ish}.
  The experimental spectral shape at $T=10$K is well described by Eq.~(1)
  with
  $n=x/5$, $\nu_{p}~=~0.5$ eV and $\gamma_{p}~=~0.3$ eV (Fig.~2). 
With the use of the polaronic and bipolaronic linewidths
  we find $\gamma\simeq 0.25$ and $\Gamma_{\rm im}\simeq 0.27$eV by taking
  for the optical phonon frequency a value $\omega=600$cm$^{-1}$
  (i.e. $\simeq   74$meV), typical 
of oxides \cite{alekor}. Then the polaron level shift is $E_p=1.0$eV
  (corresponding to the activation 
energy $E_a=125$meV), and the Coulomb energy is  $V_c=0.76$eV
in  agreement with estimates
  using the high-frequency and static dielectric constants \cite{ale},
  high-temperature dc hopping conductivity,
  phonon spectra, and the Coulomb law, respectively.

\begin{figure}[t]
  \epsfxsize=3.4in
  \epsffile{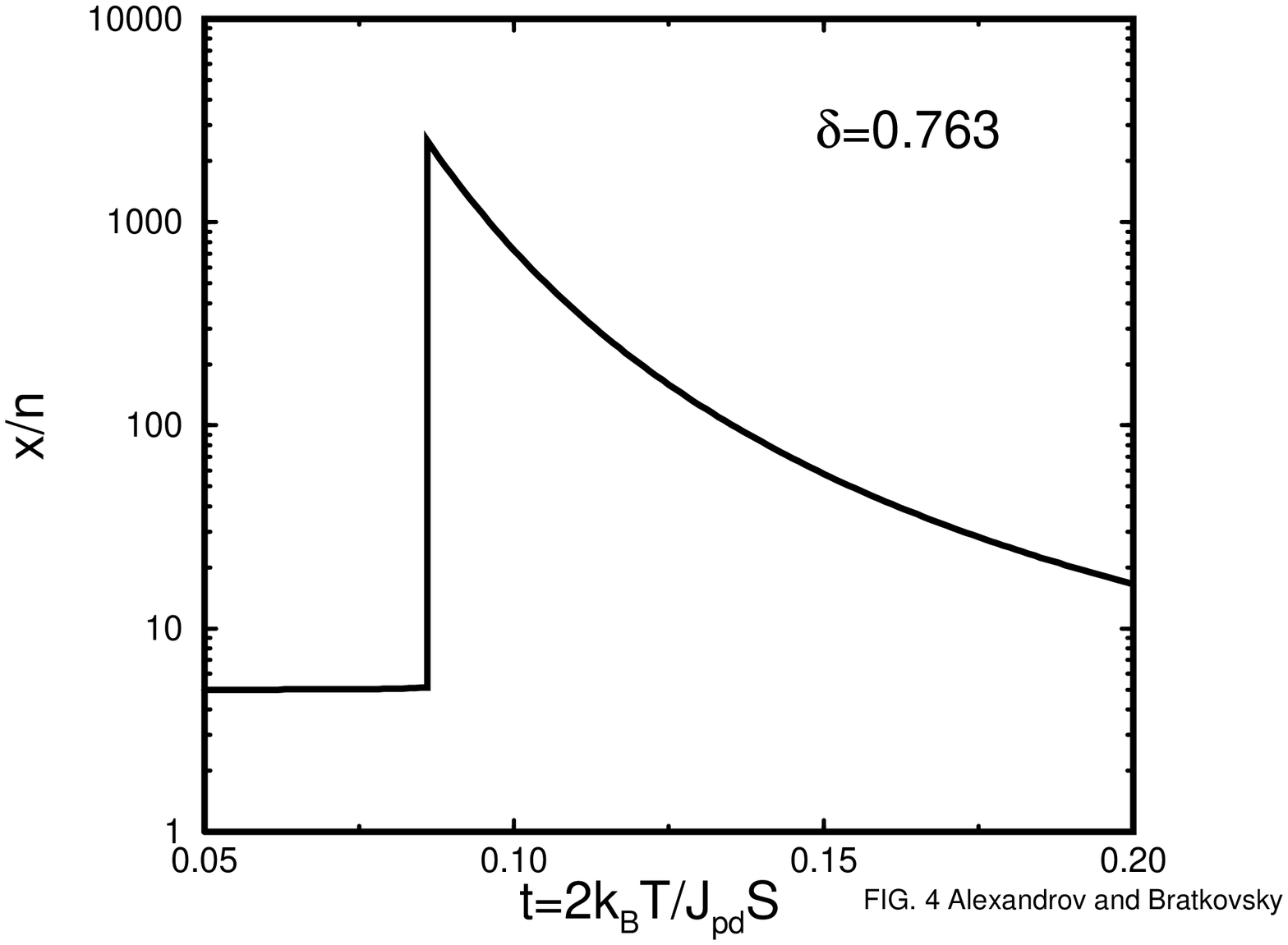 }
  \caption{Giant carrier density collapse in a charge transfer doped
  ferromagentic insulator with
  $\delta\equiv\Delta/J_{pd}S=0.763$, corresponding to the observed
  drop of dc resistivity
  below $T_c$ [15].
  \label{fig:xon}
  }
  \end{figure}

  We do not expect any significant temperature dependence of the optical
  conductivity in the paramagnetic phase because the polaron density remains
  small compared with the bipolaron density above $T_{c}$ \cite{alebra}.
  The temperature dependence of the polaron density below $T_{c}$ can be
  found
  from
  our Hartree-Fock equations \cite{alebra} generalized for
  arbitrary temperatures
  \begin{eqnarray}
  n&=& {t\over{2w}}\ln \left[{1+2y \cosh(\sigma/t) 
+ y^2\over{1+2ye^{-2w/t} \cosh(\sigma/t) + y^2e^{-4w/t}}}
  \right],\label{eq:n}\\
  m&=&{t\over{2w}}\ln \left[{1+2y e^{-w/t}\cosh({\sigma+w\over t}  ) 
+ y^2 e^{-2w/t}}\over{1+2y e^{-w/t} \cosh({\sigma-w\over{t}}) +
  y^2 e^{-2w/t}}\right],\label{eq:m}\\ 
  \sigma&=&B_2 [m/(2 t)]\label{eq:sigma},
  \end{eqnarray}
  where now
  \begin{equation}
  y= e^{-\delta/t}  
  \left[{\sinh[(x-n)d/(2xt)]\over{\sinh[(x+n)d/(2xt)]}}\right]^{1/2}.
  \end{equation}
 Here  $B_{S}(z) =\- [1+1/(2S)]\- \coth[(S+1/2)z]\- -[1/(2S)]\coth(z/2)$
  is the Brillouin function,
  $m$ and $\sigma$ are the relative magnetization of polarons and of Mn,
  respectively.
  The reduced temperature is $t=2k_{B}T/(J_{pd}S)$,
  and the dimensionless binding energy
  $\delta = \Delta/(J_{pd}S)$.  
 Compared with a nondegenerate case \cite{alebra}
 these equations take into account a finite polaron, $w =  W/J_{pd}S $
  and bipolaron, $d =  D/J_{pd}S $, widths of the energy level
  distribution, essential at low temperatures, Fig.~3. We also assume here that
  immobile bipolarons
  are localized by the impurities and there is no more than one bipolaron in
  a single localized state (`single well-single particle' approximation
  \cite{alebra3}).
  Therefore, the total number of states in the bipolaron (impurity) band is
  $x$.
 Then the bipolaron density is determined by the integral $\int
  N_{L}(E)f_{b}(E) dE$, where  $ N_{L}(E) =x/D$ the density of bipolaron
  ({\em two-particle}) impurity
  states in the energy interval $-D/2<E<D/2$, and $f_{b}(E)=
  \left[1+y^{-2}\exp[(E-\Delta)/T]\right]^{-1}$ the bipolaron distribution
  function with the chemical potential $\mu\equiv T\ln y$.
This integral should be equal to $(x-n)/2$, yielding the Eq.~(9).
 The polaron density at  zero temperature is obtained from Eq.~(6) with
  $\sigma=1$ as
\begin{equation}
n(T=0)= {2x(1-\delta)\over{d+4xw}},
\end{equation}
while in the paramagnetic phase, where $\sigma=0$ one obtains 
\begin{equation}
n(T>T_c)= {t\over{w}}\exp(-\delta/t)
\end{equation}
for $t_{c}\leq t \ll 1$. As a result there is a giant drop of the
 polaron density at $T_c$, which  
 exponentially depends on the bipolaron binding energy. Because 
 polarons are not degenerate in the
paramagnetic phase we can apply our 
nondegenerate equations \cite{alebra} to determine $t_{c}$ at a given value of $\delta$. 
We have verified that the observed 
  drop of the dc
  resistivity below
  $T_{c}$ \cite{ish}
  is well reproduced by our equations with $\delta=0.763$
 (Fig.~\ref{fig:xon}). 
The giant  drop of the polaron
density  from   $x/5$ below $T_c$ to the value $500$ times smaller above $T_c$
  is obtained with the polaron bandwidth $w\simeq 0.15$
  and with the width of the  bipolaron impurity states $d\simeq 2.25$ (for $x=0.2$).

  In conclusion, we have developed the theory
  of the optical conductivity in doped
  magnetic charge-transfer insulators with
  a strong electron-phonon interaction. We have found that the
  spectral and temperature
  features of the optical conductivity of ferromagnetic manganites are
  well described by the
  bipolaron absorption in the paramagnetic phase and by the small polaron
  absorption
  in the ferromagnetic phase. The pair breaking by exchange
  interaction with the localized Mn spins
  explains the sudden spectral weight transfer in the optical
  conductivity below $T_{c}$.
  We argue that
  the optical probe of
  the incoherent charge dynamics in manganites provides another strong
  evidence
  for the carrier density collapse which we proposed earlier as the
  explanation
  of CMR.

  We are grateful to G. Aeppli, A.R. Bishop, D.S. Dessau, M.F. Hundley,
  K.M. Krishnan, A.P. Ramirez, R.S. Williams, and Guo-meng Zhao for helpful
  discussions.
  ASA aknowledges support from the Quantum Structures Research
  Initiative and the External Research Program of
  Hewlett-Packard Laboratories (Palo Alto).


  \end{document}